\documentclass[10pt,aps,pra,twocolumn,superscriptaddress]{revtex4-2}

\usepackage[pages=all, color=black, position={current page.south}, placement=bottom, scale=1, opacity=1, vshift=5mm]{background}
\SetBgContents{
	\tt 
}      

\usepackage[margin=1in]{geometry} 

\usepackage{amsmath}
\usepackage{amsthm}
\usepackage{amssymb}
\usepackage{times}
\usepackage{braket}
\usepackage{ulem}

\usepackage{tikz}
\usetikzlibrary{quantikz2}

\usepackage[utf8]{inputenc}
\usepackage{hyperref}

\usepackage{graphicx, color}
\graphicspath{{fig/}}

\DeclareMathOperator{\E}{E}
\DeclareMathOperator{\Var}{Var}
\DeclareMathOperator{\Cov}{Cov}
\DeclareMathOperator{\Perm}{Perm}

\usepackage{algorithm, algpseudocode} 
\usepackage{mathrsfs} 
\usepackage{lipsum}

\begin{document}

\title{Efficient classical algorithm for simulating boson sampling with heterogeneous partial distinguishability}

\author{S.N. van den Hoven}
\affiliation{MESA+ Institute for Nanotechnology, University of Twente, P.~O.~box 217, 7500 AE Enschede, The Netherlands} 

\author{E. Kanis}
\affiliation{MESA+ Institute for Nanotechnology, University of Twente, P.~O.~box 217, 7500 AE Enschede, The Netherlands} 

\author{J. J. Renema}
\affiliation{MESA+ Institute for Nanotechnology, University of Twente, P.~O.~box 217, 7500 AE Enschede, The Netherlands} 

\date{
	\today
}

\begin{abstract}
Boson sampling is one of the leading protocols for demonstrating a quantum advantage, but the theory of how this protocol responds to noise is still incomplete. We extend the theory of classical simulation of boson sampling with partial distinguishability to the case where the degree of indistinguishability between photon pairs is different between different pairs. 
\end{abstract}

\maketitle

\section{Introduction}
\label{sec:intro}
Quantum computers are expected to outperform classical computers in certain well-defined tasks such as the hidden subgroup problems for abelian finite groups, which includes prime factorization \cite{Shor_prime}, simulations of quantum systems\cite{seth_lloyd_simulating_quantum_sys}, and unstructured search \cite{Grovers_algorithm}.  However, building a universal fault-tolerant quantum computer is no easy task, due to the extreme degree of control over a large number of quantum particles required. As an intermediate step, experimental research has focused on the demonstration of a quantum advantage \cite{harrow2017}, i.e. a computational task where quantum hardware outperforms all classical hardware in wall-clock time, on a well-defined computational problem not necessarily of any practical utility. Such demonstrations have been claimed in superconducting circuits \cite{arute2019quantum, Wu2021, Zhu2022}, and photons \cite{zhong2020quantum, wang2019, zhong2021}. 

These quantum advantage claims caused substantial debate, with several later being outperformed by classical simulations \cite{Pednault2019, PanZhang2021, Tindall2024, Oh2023_1, Sigbovik2024}. This was possible despite strong guarantees of computational complexity because experimental hardware suffers from various forms of noise, which introduce decoherence, and reduce the degree to which the task which the device is performing is truly quantum mechanical, thereby opening up loopholes for classical simulation strategies to exploit. Similar to the situation in Bell tests, these simulation strategies demarcate the regime where a classical explanation for the observed data cannot be ruled out. They therefore serve a vital function in assessing the success or failure of a quantum advantage demonstration.  

\begin{figure}
    \centering
    \includegraphics[width=0.5 \textwidth]{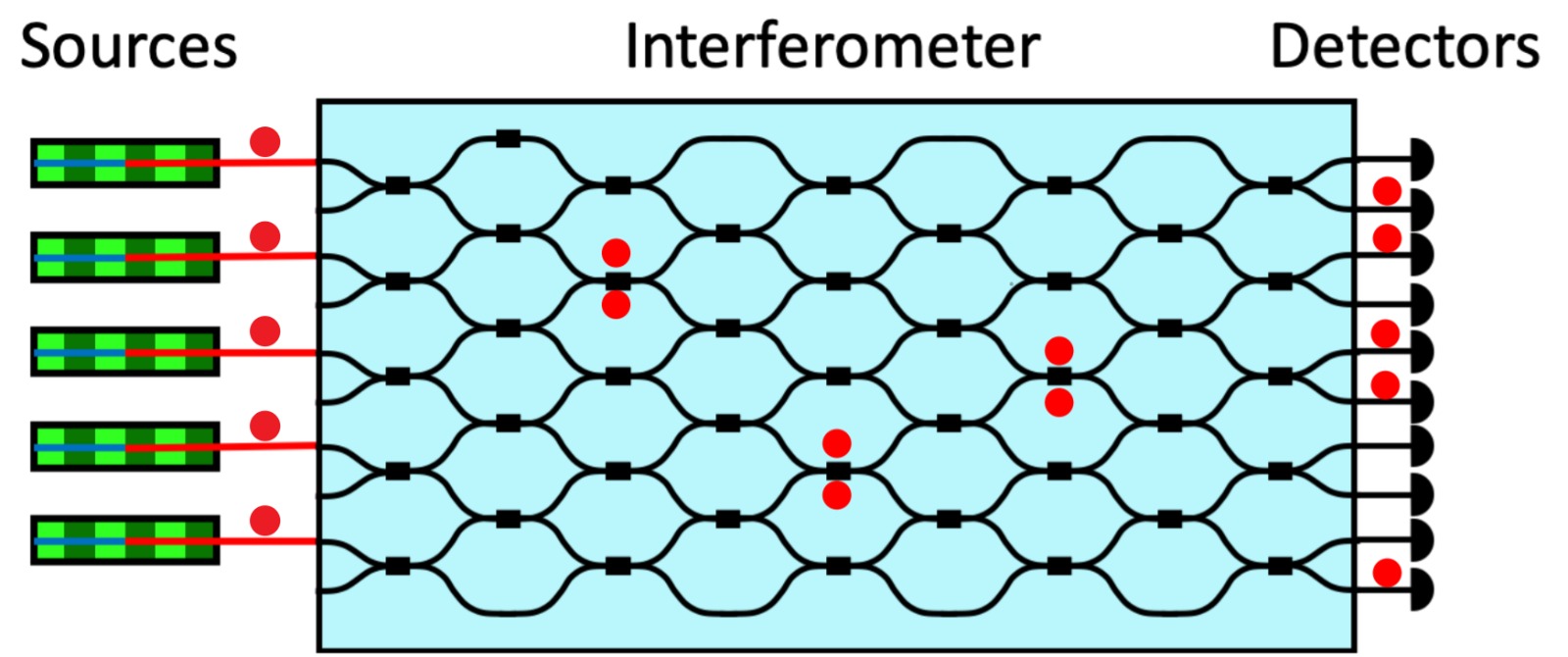}
    \caption{Schematic representation of a boson sampling device. The photon states
enter the linear optical network on the left. The photons interfere and create an output
state. Upon measuring where the photons leave the interferometer, a sample from the
created output state has been drawn}
    \label{fig:schem boson sample}
\end{figure}

One protocol for a quantum advantage demonstration is boson sampling \cite{Aaronson_Arkhipov_BS}. In boson sampling, single photons are sent through a large-scale linear interferometer. The computational task is to provide samples from the output state measured in the Fock basis (see Fig. \ref{fig:schem boson sample}). Complexity arises ultimately from quantum interference between the exponentially many ways in which the photons can traverse the interferometer to produce a single outcome. The main sources of noise in boson sampling are photon loss \cite{brod2020classical}, where some of the photons do not emerge from the output of the interferometer, and photon distinguishability \cite{tichy2015sampling}, where the particles carry which-path information in their internal quantum states. 

Several strategies exist to classically simulate imperfect boson sampling, including ones based on approximating the quantum state using tensor networks \cite{Lubasch2018, Oh2023_1, Cilluffo2023, Liu2023}, ones aimed at reproducing the marginal photon distributions behind some number of optical modes \cite{Villalonga2021, Bulmer2022}, and ones based on based on phase-space methods \cite{rahimikeshari2016,  Qi2020, Drummond2021, Dellios2023}. Some methods are specifically aimed at spoofing certain benchmarks which have themselves been put forward as proxies for computational complexity \cite{Oh2023_3}. 

The classical simulation technique that we focus on here makes use of the fact that imperfections dampen quantum interference more strongly between paths through the interferometer that exhibit a higher degree of classical dissimilarity, i.e. which would be more different if the particles were fully classical \cite{renema2018efficient, renemaarxiv2018}. This allows us to establish a notion of distance between the paths, with the attenuation of quantum interference between two paths depending exponentially on the distance. Since it can be shown that there are only polynomially many paths shorter than a given distance, truncating the quantum interference at a fixed distance produces an approximation to the output probability, which is both efficiently computable and maintains its accuracy as the system size is scaled up. 

Interestingly, this bosonic algorithm has a direct counterpart in the simulation of qubit-based systems \cite{Aharonov2023}, as do some of the other algorithms. It is an open question whether this is a coincidence or a symptom of some deeper structure of the problem of demonstrating a quantum advantage, with Kalai and Kindler conjecturing \cite{KalaiKindler2014} that the susceptibility of boson sampling to noise is an intrinsic feature of a non-error corrected approach to demonstrating a quantum advantage. 

However, this algorithm \cite{renema2018efficient} suffers from some restrictions. In particular, it assumes that the degree of indistinguishability between all photons is equal. In the case of varying indistinguishability among pairs of photons, the only solution available to the algorithm is to approximate the degree of indistinguishability between all photons as that of the highest pairwise indistinguishability. This substantially reduces the applicability of the algorithm to a real experiment, where such fluctuations inevitably occur. In the most extrema case, an experiment with two fully indistinguishable photons and otherwise all distinguishable photons could not be classically simulated, even though only two-photon quantum interference occurs in this case.

In this work, we eliminate the dependency on this assumption, demonstrating a classical simulation of noisy boson sampling that is efficient for realistic models of dissimilar photon indistinguishability. We focus on the case where the partial distinguishability is modeled according to a generalized version of the orthogonal bad-bit model \cite{sparrow2018phd_thesis,marshall2022distillation}. The results of \cite{renema2018efficient} follow as a special case of this model. Furthermore, this model allows us to deal with the extreme case in which two photons are indistinguishable, while the others are completely distinguishable. We show an extension of the algorithm of \cite{renema2018efficient} which generally achieves better performance than the original, extending the area of the parameter space which is susceptible to classical simulation.
We find that the complexity is entirely governed by the quadratic mean of the distribution of Hong-Ou-Mandel \cite{hong1987measurement} visibilities. 
Hence, we find that heterogeneity in the partial distinguishability makes simulation more expensive than the homogeneous case with equal arithmetic means of partial distinguishability, but not prohibitively so.

We achieve these results by reworking and simplifying the derivation of \cite{renema2018efficient}, to more easily accommodate more complex partial distinguishability distributions. We therefore show that the sensitivity of the hardness of boson sampling to imperfections is not a result of the specific assumptions made in the classical simulation techniques. Moreover, these results provide evidence for the idea that the sensitivity of quantum advantage demonstrations to noise is intrinsic rather than dependent on the specific model of noise chosen.

We focus specifically on partial distinguishability as a source of error, motivated by the idea that both optical loss and indistinguishability, as well as other errors, all affect the computational complexity of the boson sampling problem in similar ways \cite{renemaarxiv2018,renema2021sample}, meaning that any is paradigmatic for the others. We leave full extension of our results to optical loss and other imperfections to future work.

\section{Classical algorithm for boson sampling with partially distinguishable photons}
\label{sec:easy proof}
In this section, we will revisit the algorithm for efficiently classically simulating boson sampling with partial distinguishable particles as described in \cite{renema2018efficient, GarciaPatron2019simulatingboson}. We demonstrate a simplified proof for the algorithm, that is heavily inspired by the proof given in \cite{renema2018efficient}, but allows us to extend the algorithm more easily. 

This section will be structured as follows. First we will revisit the theory needed to describe interference experiments with partial distinguishable photons. Then, we will show that we can efficiently approximate transition probabilties by neglecting contributions of high-order multiphoton interference. Lastly we will show that the error induced by such an approximation on the total probability distribution is independent of the number of photons.

We start by considering boson sampling with partially distinguishable photons. 
Previous research has demonstrated a method to compute the probability of detecting a certain output configuration \textbf{s} in the Fock basis \cite{shchesnovich2014sufficient,shchesnovich2015partial}.
This expression allows for arbitrary multi-photon input states and arbitrary number-resolving photon detectors. 

Under the assumption of pure input states and lossless detectors which are insensitive to the internal state of the photon, it has been shown that the results of \cite{shchesnovich2014sufficient,shchesnovich2015partial} can be rewritten in the a compact form \cite{tichy2015sampling}. The probability of measuring a particular detection outcome \textbf{s} is given by:
\begin{equation}
    P=\frac{1}{\prod_i r_i! s_i!} \sum_{\sigma \in S_n} \left [ \prod^n_{j=1} \mathcal{S}_{j,\sigma(j)}\right ] \mathrm{perm}(M \circ M^{*}_{\sigma}),
    \label{eq:Tichy,multiperm}
\end{equation}
where $M$ is a submatrix of the unitary representation of the interferometer $U$, constructed by selecting rows and columns of $U$ corresponding to the input modes and output modes of interest ($M=U_{d(\textbf{r}),d(\textbf{s})}$), where $d(\textbf{r})$ and $d(\textbf{s})$ represent the mode assignment lists of the input and output states respectively. Note that the size of $M$ is determined by the number of photons $n$ considered in the sampling task. $r_i$ and $s_i$ denote the $i^{th}$ element of the mode occupation lists of the input state and the output state respectively. $S_n$ denotes the symmetric group, $\mathcal{S}$ denotes the distinguishability matrix where $\mathcal{S}_{i,j}=\langle \psi_i|\psi_j\rangle$ denotes the overlap between photon $i$ and $j$ represented by their internal wave-functions $\psi_i$ and $\psi_j$ respectively.  $\circ$ represents the Hadamard product, $^*$ denotes the element wise conjugation and $M_{\sigma}$ denotes $M$ where its rows are permuted according to $\sigma$. Lastly the permanent of matrix $M$ with shape $n \times n$ is defined as 
\begin{equation}
        \text{Perm}(M) = \sum_{\sigma \in S_n}\prod_{i=1}^n M_{i, \sigma(i)}.
\end{equation}

For the moment, we will continue by assuming that all particles are equally distinguishable, i.e. that: 
\begin{equation}
        \mathcal{S}_{ij}=x+(1-x)\delta_{ij}.
    \label{eq: sij=x+(1-x)}
\end{equation}
Note that the assumption made in Eq. (\ref{eq: sij=x+(1-x)}) is in the literature often referred to as the orthogonal bad-bit model \cite{sparrow2018phd_thesis, shaw2023errors}, and is supported by experimental evidence. 

Using the assumption of Eq. (\ref{eq: sij=x+(1-x)}), we note that
the quantity $\prod^n_{j=1} \mathcal{S}_{j,\sigma(j)}$ will only depend on the number of fixed points of $\sigma$. (A fixed point is a point in $\sigma$ the that maps to itself after the permutation, or $\sigma(j)=j$.) We can therefore rewrite Eq. (\ref{eq:Tichy,multiperm}) as: 
\begin{equation}
    P(s)=\frac{1}{\prod_i r_i! s_i!}\sum_{j=0}^n\sum_{\tau \in \sigma_j}x^j \text{Perm}(M \circ M^*_{\tau}).
    \label{eq: distinguishable probability with S=x}
\end{equation}
Here, $\sigma_j$ denotes the set of all permutations with $n-j$ fixed points. 

The term $x^j$ introduces exponential dampening in $j$. For this reason, it is natural to truncate the series at some value $k<n$:
\begin{equation}
    P_k(s)=\frac{1}{\prod_i r_i! s_i!}\sum_{j=0}^k\sum_{\tau \in \sigma_j}x^j \text{Perm}(M \circ M^*_{\tau}),
    \label{eq: truncated distinguishable probability with S=x}
\end{equation}
leaving an expression for the error:
\begin{equation}
    Q_k=\frac{1}{\prod_i r_i! s_i!}\sum_{j=k+1}^n\sum_{\tau \in \sigma_j}x^j \text{Perm}(M \circ M^*_{\tau}).
    \label{eq: error truncation distinguishable probability with S=x}
\end{equation}

Note that all terms in Eq. (\ref{eq: distinguishable probability with S=x}) for a given $j$ correspond to all contributions to the probability where $j$ photons interfere with each other and $n-j$ photons undergo classical transmission. Hence, by truncating Eq. (\ref{eq: distinguishable probability with S=x}), we only consider those contributions to the probability where at most $k$ photons interfere with each other.

In the remainder of this section, we will focus on non-collisional input and output states, hence $\prod_i (r_i)! (s_i)!=1$. We will show two things. The first is that Eq. (\ref{eq: truncated distinguishable probability with S=x}) can be computed efficiently on a classical computer for arbitrary system sizes $n$. The second is that, under the assumption that $M$ is filled with elements that are i.i.d. complex Gaussian, the error term in Eq. (\ref{eq: error truncation distinguishable probability with S=x}) decreases exponentially as $n$ increases. It decreases such that, the expectation value of the $L_1$-distance between the approximate distribution and the real distribution is upper bounded by the following expression:
\begin{equation}
    \E\left(\sum_s |P(s)-P_k(s)|\right)<\sqrt{\frac{x^{2k+2}}{1-x^2}}.
    \label{eq: upper bound}
\end{equation}
This upper bound holds regardless of the system size.
Note that if the number of modes is much larger than the number of photons, any $n\times n$ submatrix of a Haar-random matrix will be close in variation distance to a matrix filled with complex i.i.d. Gaussians \cite{Aaronson_Arkhipov_BS}.
We can use the first result to efficiently draw samples from the approximate distribution via a Metropolis sampler, and we use the second result to show that the distribution that is sampled from is close in variation distance to the real distribution, thus resulting in an efficient classical algorithm for imperfect boson sampling. 

To demonstrate that we can efficiently compute $P_k$ from Eq. (\ref{eq: truncated distinguishable probability with S=x}), we use the fact that an algorithm exists for approximating the permanent of a matrix with real non-negative elements \cite{JSV_algorithm}. Additionally, we use the Laplace expansion to split up the permanents from Eq. (\ref{eq: error truncation distinguishable probability with S=x}) into sums of the product of two permanents of smaller matrices. One of these two permanents is filled with non-negative elements.
We rewrite each term in Eq. (\ref{eq: truncated distinguishable probability with S=x}) by Laplace expanding
about the rows that correspond to the fixed points of $\tau$:
\begin{multline}
        \text{Perm} (M \circ M^*_{\tau})=\\ \sum_{\rho \in \big(\begin{smallmatrix}
n \\
j 
\end{smallmatrix}\big)} \text{Perm} (M_{I_{p},\rho} \circ M^*_{\tau_{p},\rho})\text{Perm}( |M_{\tau_u,\bar{\rho}}|^2).
    \label{eq: laplace expand perm M M}
\end{multline}
Here $\tau_u$ and $\tau_p$ are the unpermuted and permuted parts of $\tau$ respectively, i.e. those parts that correspond to fixed points (cycles of length 1) and longer cycles, respectively. Given that $\tau$ has $n-j$ fixed points, $\rho$ is a $j$-combination of $n$, $\bar{\rho}$ is its complementary set and $I_{p}$ is the identity permutation for the elements of $\rho$. 
Eq. (\ref{eq: laplace expand perm M M}) now contains two permanents. The second permanent contains a matrix with only real non-negative elements and can be efficiently approximated via the JSV algorithm \cite{JSV_algorithm}. The other permanent contains a matrix with complex elements. The size of these matrices is however determined by $j$, which due to our truncation has a maximum value of $k$.
To compute Eq. (\ref{eq: truncated distinguishable probability with S=x}) we thus need to compute permanents of complex matrices of size $j$ and permanents of real, non-negative matrices of size $n-j$. We need to do both of these calculations $\big(\begin{smallmatrix}
n \\
j 
\end{smallmatrix}\big)$ times for each $\tau\in\sigma_j$ for all $j\leq k$, which results in a polynomial scaling of computational costs with $n$ to evaluate $P_k$. 

We now continue with a derivation for Eq. (\ref{eq: upper bound}). In the main text we will give a sketch of this derivation, in Appendix \ref{sec: derivation upper bound} we will give a full derivation.\footnote{The derivation in Appendix \ref{sec: derivation upper bound} is given for the more general situation as discussed in section \ref{sec: general part dist}. To consider the assumption made in Eq. (\ref{eq: sij=x+(1-x)}), we simply consider the special case where $x_i=x$ for all $i$. Then $\sqrt{\mathrm{M}_2}=x^2$ and Eq. (\ref{eq: upper bound}) is equal to Eq. (\ref{eq: upper bound appendix}) } It is important to note that the derivation follows the same ideas as presented in \cite{renema2018efficient}, but differs in some key details. These differences allow us to find similar upper bounds for adjacent boson sampling experiments as will be elaborated on in the following sections. 

The derivation consists of the following steps:
\begin{enumerate}
    \item Using Jensen's inequality, note that $\E(|Q_k|)\leq \sqrt{\Var(Q_k)}$
    \item Using the definition of the permanent, note that $Q_k$ is a large sum where each term is described by a product containing elements of $M$
    \item Using Bienamaymé's identity, note that the variance of a large sum is equal to the covariance between all pairs of terms in this large sum
    \item $M$ is assumed to be filled with i.i.d. complex Gaussian elements, and hence $\E(M_{ij})=0$, $\E((M_{ij})^2)=0$, $\E((M^{*}_{ij})^2)=0$, $\E(|M_{ij}|^2)=\frac{1}{m}$ and $\E(|M_{ij}|^4)=\frac{2}{m^2}$ for all $i,j$.
    \item We use these properties of the elements of $M$ to find that almost all of these correlations are equal to zero.
    \item We use simple combinatorics to count the number of covariances that contribute the same non-zero amount to $\Var(Q_k)$
    \item We find that the error term approximates a truncated geometric series
    \item We have now found an approximate expression for $\Var(Q_k)\approx \frac{(n!)^2}{m^n}\frac{x^{2k+2}-x^{2n+2}}{1-x^2}$ and as a result an upper bound for $\E(|Q_k|)$ for a typical non-collisional output configuration
    \item By counting the number of non-collisional output configurations we find the upper bound on $L_1$-distance of interest as presented in Eq. (\ref{eq: upper bound})
\end{enumerate}
To conclude this section, we have revisited a known algorithm as described in \cite{renema2018efficient, GarciaPatron2019simulatingboson}. The algorithm approximates transition probabilities in a boson sampling experiment by neglecting high order interference contributions to the transition probability. We have demonstrated that these approximated transition probabilities are efficiently computable. Moreover we have demonstrated that, under the assumption that all photons are equally distinguishable, the $L_1$-distance over all non-collisional outputs between the approximated distribution and the real distribution is upper bounded as demonstrated in Eq. (\ref{eq: upper bound}). Notably, this upper bound is independent of the system size of the boson sampling experiment. In the following section we will relax this assumption.

\section{General partial distinguishability}
\label{sec: general part dist}
In the previous section, we have sketched an efficient classical algorithm for boson sampling with partially distinguishable photons, with full details given in Appendix A. To show that the approximate distribution we can efficiently sample from is close to the real distribution it was assumed that all particles are equally distinguishable, see Eq. (\ref{eq: sij=x+(1-x)}). 
However, realistic single-photon sources do not adhere to the assumption that all photons are equally imperfect. The algorithm as proposed in \cite{renema2018efficient} circumvents this problem by computing the upper bound in Eq. (\ref{eq: upper bound}) as if all photons are equally imperfect and
as good as the best photon pair present. This way an upper bound can be found in general, but depending on the variations in the quality of the particles, this bound may be very loose.
Here, we relax the assumption made in Eq. (\ref{eq: sij=x+(1-x)}).
Although we are not able to tighten this upper bound in general, we show that this bound can be tightened under the assumption that the partial distinguishability of the photons is modeled according to a generalized orthogonal bad-bit model. In this generalization, the HOM-visibility may vary for two different pairs of photons. We argue that this generalization is experimentally relevant for two reasons. First, it covers the extreme case in which two photons are indistinguishable, while the others are completely distinguishable. 
Second, upon manufacturing single-photon sources, the target mode is most likely the same for all sources but deviations in the manufacturing process result in fluctuations in the quality of the individual sources. The target mode is captured in the first term and the imperfections are captured in the second term of Eq. \ref{eq: iid obb}.
\footnote{A more realistic model would allow for the possibility of interference between parts of the wavefunction that are not covered by the target mode. In other words, $\langle\Psi_i|\Psi_j\rangle=\delta_{ij}$ in Eq. (\ref{eq: iid obb}) would no longer be true. We believe that this additional interference that is not covered by our model would add little complexity since in realistic scenario's $|\sqrt{x_i x_j}|^2>>|\sqrt{(1-x_i)(1-x_j)}\langle\Psi_i|\Psi_j\rangle|^2$.}

We will start by noting that without the assumption made in Eq. (\ref{eq: sij=x+(1-x)}), we can still efficiently evaluate our truncated probability. For general, pure \footnote{As long as our input state is described by a mixture of polynomially many pure states, this algorithm can be used to compute the cumulative probability of measuring a particular detection outcome of interest efficiently. If the mixture contains exponentially many pure states, an approximation is needed that reduces this mixed state to a mixed state with polyonomially many terms} partially distinguishable photons, Eq. (\ref{eq: truncated distinguishable probability with S=x}) reads:

\begin{equation}
    P_k=\frac{1}{\prod_i r_i! s_i!}\sum_{j=0}^k\sum_{\tau \in \sigma_j}\left(\prod^n_{i=1}\mathcal{S}_{i,\tau(i)}\right) \text{Perm}(M \circ M^*_{\tau}).
    \label{eq: truncated distinguishable probability with general S}
\end{equation}
If we compare Eq. (\ref{eq: truncated distinguishable probability with general S}) with Eq. (\ref{eq: truncated distinguishable probability with S=x}), we notice that the only difference is that $x^j$ is substituted with $\prod^n_{i=1}\mathcal{S}_{i,\tau(i)}$. It takes a multiplication of $j$ factors to compute $\prod^n_{i=1}\mathcal{S}_{i,\tau(i)}$ and hence Eq. (\ref{eq: truncated distinguishable probability with general S}) is still efficiently computable on a classical computer.

We will continue to derive an upper bound for the $L_1$-distance between the approximate and the real distribution.
For general, pure partially distinguishable photons, Eq. (\ref{eq: error truncation distinguishable probability with S=x}) becomes
\begin{equation}
    Q_k=\frac{1}{\prod_i r_i! s_i!}\sum_{j=k+1}^n\sum_{\tau \in \sigma_j}\left(\prod^n_{i=1}\mathcal{S}_{i,\tau(i)}\right) \text{Perm}(M \circ M^*_{\tau}).
    \label{eq: error truncation distinguishable probability with General S}
\end{equation}

We note that $\prod^n_{i=1}\mathcal{S}_{i,\tau(i)}$ is completely independent of all elements in $M$. After all, the choice for the interferometer is completely independent of the quality of the photons used in the experiment. We are considering the expectation value of $|Q_k|$ over the ensemble of Haar-unitaries, we thus note that $\E\left(\prod^n_{i=1}\mathcal{S}_{i,\tau(i)}\right)=\prod^n_{i=1}\mathcal{S}_{i,\tau(i)}$. As a result, the same sketch of the derivation as used in section \ref{sec:easy proof} applies and again for the full derivation, we refer to Appendix \ref{sec: derivation upper bound}. 
For any partial distinguishability model, all steps in Appendix \ref{sec: derivation upper bound} are valid up until Eq. (\ref{eq: var (Q) intermediata}): 
\begin{multline}
    \Var(Q_k)=\\
    =\sum_{j=k+1}^n \sum_{\tau\in\sigma_j}\prod^n_{i=1}|\mathcal{S}_{i,\tau(i)}|^2 n!\sum^{n-j}_{p=0}\mathrm{R}_{n-j,p}2^p\left(\frac{1}{m^2}\right)^{n}
    \label{eq: var intermediate main}
\end{multline}

We would like to find an expression (or an upper bound) for $\prod^n_{i=1}|\mathcal{S}_{i,\tau(i)}|^2$ that is independent of $\tau$, because that would allow us to simplify Eq. (\ref{eq: var intermediate main}) further by recognizing a truncated geometric series. We note that we can always simplify Eq. (\ref{eq: var intermediate main}) by realizing that $\prod^n_{i=1}|\mathcal{S}_{i,\tau(i)}|^2\leq \mathrm{max}(|\mathcal{S}_{ij}|)^{2j}$, this approach has been mentioned in \cite{renema2018efficient} but could be very costly. In general, we believe that it is difficult to tighten the bound resulting from the above-mentioned approach, because the structure of $\sum_{\tau\in\sigma_j}\prod^n_{i=1}|\mathcal{S}_{i,\tau(i)}|^2$ changes significantly for different values of $j$. 

In the remainder of this section, we will therefore restrict ourselves to a generalized version of the orthogonal bad-bit model \cite{sparrow2018phd_thesis,marshall2022distillation}. For this model, we are able to tighten the upper bound for the $L_1$-distance in Eq. (\ref{eq: var intermediate main}).

For the generalized version of the orthogonal bad-bit model \cite{sparrow2018phd_thesis,marshall2022distillation}, we will consider the internal modes of the $i^{th}$ photon to be:
\begin{equation}
    |\psi\rangle = \sqrt{x_i}|\Psi_0\rangle + \sqrt{1-x_i}|\Psi_i\rangle,
    \label{eq: iid obb}
\end{equation}
where $\langle\Psi_i|\Psi_j\rangle=\delta_{ij}$. In this generalization, we allow for deviations in the quality of the photons. With this assumption, we can simplify Eq. (\ref{eq: var intermediate main}).

The overlap matrix $\mathcal{S}$ for this model becomes:
\begin{equation}
  \mathcal{S}_{ij} =
    \begin{cases}
      1  & \text{for $i=j$}\\
      \sqrt{x_i}\sqrt{x_j}^* & \text{for $i\neq j$}
    \end{cases}       
\end{equation}
To clarify how this model simplifies Eq. (\ref{eq: var intermediate main}), we realize that every permutation $\tau$ can uniquely be described with its cycle notation. From the cycle notation it becomes clear that

\begin{equation}    \sum_{\tau\in\sigma_j}\prod^n_{i=1}|\mathcal{S}_{i,\tau(i)}|^2=\mathrm{R}_{n,n-j} \mathrm{M}_j.
    \label{eq: E[prod S]=prod E[x^2]}
\end{equation}
Here $\mathrm{R}_{n,n-j}$ is rencontres number that counts the number of ways one can permute the set $\{1,\cdots, n\}$ with $n-j$ fixed points. $\mathrm{M}_j$ is the elementary symmetric polynomial of order $j$ for the variables $|x_i|^2$, divided by $\begin{pmatrix}n\\j\end{pmatrix}$.
We can use McLaurin's inequality: \footnote{McLaurin's inequality can be understood as a refinement on the inequality between arithmetic and geometric means.}
\begin{equation}
\begin{split}
    \sum_{\tau\in\sigma_j}\prod^n_{i=1}|\mathcal{S}_{i,\tau(i)}|^2&=\mathrm{R}_{n,n-j} \mathrm{M}_j\\
    &\leq \mathrm{R}_{n,n-j} \sqrt{\mathrm{M}_2}^j
    \end{split}
    \label{eq: mclaurin's inequality}
\end{equation}
Where $\sqrt{\mathrm{M}_2}$ represents the quadratic mean of $|x_i x_j|$. Which can more conveniently be recognized as the quadratic mean of all pairwise HOM-visibilities \cite{hong1987measurement} \footnote{We could also express our final bound in terms of $M_1$, which is the arithmetic mean of $|x_i|^2$. This would lead to a tighter bound. However, we choose to describe our final result in terms of $M_2$ to clarify the link between the upper bound and the well-known HOM visibilities.}.

Using McLaurin's inequality from Eq. (\ref{eq: mclaurin's inequality}), we managed to find an upper bound for Eq. (\ref{eq: var intermediate main}) which is independent of $\tau$ and which describes a truncated geometric series. Further details of the simplification of this bound can be found in Appendix \ref{sec: derivation upper bound}.

After these simplifications we find the following expression for an upper bound on the $L_1-$distance:

\begin{align}
    \E\left(\sum_s|P(s)-P_k(s)|\right)&=\E\left(\sum_s|Q_k(s)|\right)\\
    &<\sqrt{\frac{( \sqrt{\mathrm{M}_2}^{k+1}}{1- \sqrt{\mathrm{M}_2}}}.
    \label{eq: upperbound general S main text}
\end{align}

We find that Eq. (\ref{eq: upperbound general S main text}) is equal to Eq. (\ref{eq: upper bound}), with the substitution of $\sqrt{\mathrm{M}_2}$ for $x^2$. For sources that follow the generalized orthogonal bad-bit model, the complexity of boson sampling is therefore governed by the quadratic mean of the HOM visibilities. This result improves on earlier work, which could only upper bound the complexity of boson sampling on general sources by the maximum of their visibilities. 

We continue by comparing this newly found upper bound to the previous upper bound that one would get while using the maximum pairwise distinguishability in Eq. \ref{eq: upper bound}. 
For this comparison, we consider the specific case in which all $|x_i|$ from Eq. (\ref{eq: iid obb}) are considered independent and identically distributed random variables, distributed according to a Gaussian distribution with mean $\mu$ and standard deviation $\sigma$. \footnote{A realistic scenario in which an experimentalist would like to defend his/her claim to a quantum advantage, the experimentalist should know the quality of the photons. The experimentalist could use these qualities to figure out whether this algorithm could efficiently simulate the boson sampling experiment given an acceptable error margin.}
Form McLaurin's inequality we know $\sqrt{M_2}<M_1$ and in the limit of $n \rightarrow \infty$, $M_1=\E\left(|x_i|^2\right)$. For $|x_i|\sim \mathcal{N}(\mu,\sigma^2)$ we find $\E(|x_i|^2)=\mu^2+\sigma^2$.

\begin{figure}[h!]
    \centering
    \includegraphics[width=0.45\textwidth]{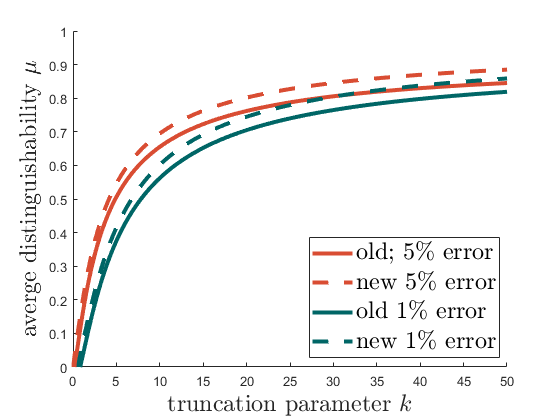}
    \caption{Solutions to Eqs. (\ref{eq: upper bound}) (solid) and (\ref{eq: upperbound general S main text}) (dashed) for two different upper bounds on the $L_1$-distance ($1\%$ and $5\%$).
    We consider an i.i.d. orthogonal bad-bit model where the standard deviation is given by $\sigma=0.02$. Eq. (\ref{eq: upper bound}) assumes $\mathrm{max}(x_i)=\mu+2 \sigma$. The lines indicate pairs of $\mu$ and $k$ for which an upper bound on the $L_1$-distance is given by $1\%$ (blue) or $5\%$ (orange).
    }
    \label{fig2: iid obb}
\end{figure}

Figure \ref{fig2: iid obb} shows the effect of this tighter bound on the regime of the parameter space which can be efficiently simulated. We assume $\mathrm{max}(x_i)=\mu+2 \sigma$, motivated by the observation that  the probability is then about $\frac{1}{2}$ that $\mathrm{max}(x_i)<\mu+2\sigma$ in the case of 30 photons already, and decreasing further if $n$ increases. The areas of the parameter space in Fig \ref{fig2: iid obb} which are in between the solid and dotted lines are the areas of the parameter space which could not be simulated before.

\section{Discussion and conclusion}
\label{sec: discussion and conclusion}
We have extended classical simulation techniques for noisy boson sampling. These new classical simulation techniques push the boundaries for the required qualities of the resources needed for an optical quantum computer. Our results strengthen the intuition that the fragility of computational complexity to noise is itself a robust phenomenon, that does not depend on the particular details of how that noise is modeled. Future work will focus on including other heterogeneous noise sources into this model, such as unbalanced optical loss.  

\section{Acknowledgments}
\label{sec: Acknowledgments}
We thank F.H.B Somhorst, M. Correa Anguita and C. Toebes for scientific discussions. This research is supported by the PhotonDelta National Growth Fund program. This publication is part of the project At the Quantum Edge (VI. Vidi.223.075) of the research programme VIDI which is financed by the Dutch Research Council (NWO).

\appendix
\section{Formal derivation for the upper bound on the $L_1$-distance}
\label{sec: derivation upper bound}
In this section we will give a formal derivation of the upper bound as presented in
Eqs. (\ref{eq: upper bound}) and (\ref{eq: upperbound general S main text}).
We will follow the steps as presented in the main text. In this derivation we will assume that the transition probability amplitudes can be considered as elements, sampled from an i.i.d. complex Gaussian distribution and that the internal wavefunctions of our particles follow a generalized orthogonal bad-bit model. We clearly state when we use these assumptions. \\

We start by realizing that $Q_k$ from Eq. (\ref{eq: error truncation distinguishable probability with General S}) is real. This follows from the fact that 
\begin{multline}
 \left(\prod^n_{i=1}\mathcal{S}_{i,\tau(i)}\right) \text{Perm}(M \circ M^*_{\tau}) =\\
 \left(\left(\prod^n_{i=1}\mathcal{S}_{i,\pi(i)}\right) \text{Perm}(M \circ M^*_{\pi})\right)^*,   
\end{multline}
Which is true for $\pi = \tau^{-1}$.Therefore we know $|Q_k|=\sqrt{Q_k^2}$. 
Jensen's inequality for a concave function \cite{dekking_computations_2005} yields
\begin{equation}
\label{eqn:E_Q_jensen}
    \E(|Q_k|)=\E(\sqrt{Q_k^2})\leq \sqrt{\E(Q_k^2)} 
\end{equation}
We note
\begin{equation}
\begin{split}
        \E(\Perm(M\circ M_{\tau}^*))& = \\
    \E(\sum_{\rho \in S_n}\prod_{i=1}^n(M\circ M_{\tau}^*)_{i,\rho(i)})& =\\
    \sum_{\rho \in S_n}\E(\prod_{i=1}^n(M\circ M_{\tau}^*)_{i,\rho(i)})&,
    \label{E Perm M M=0}
\end{split}
\end{equation}

where we used the definition of the permanent and the linearity of the expectation value.
We are interested in the situation where our truncation parameter $k$ is larger than zero, and as a result, all permutations $\tau$ that we consider have a nonzero amount of points which are not fixed. In other words:
\begin{equation}
    \exists \: q \: \mathrm{s.t}.\: \tau(q)\neq q \: \forall \: \tau
    \label{eq: at least 1 nonfixed point}
\end{equation}
and Eq. (\ref{E Perm M M=0}) can be written as
\begin{multline}
        \sum_{\rho \in S_n}\E(\prod_{i=1}^n(M\circ M_{\tau}^*)_{i,\rho(i)}) =\\
    \sum_{\rho \in S_n}\E \left( \prod_{\substack{i=1\\i\neq q}}^n(M\circ M_{\tau}^*)_{i,\rho(i)}\right)\E(M_{q,\rho(q)}) \E(M_{\tau(q),\rho(q)}^*) \\= 0.
    \label{eq: exp Q=0 because at least 1 nonfixed point}
\end{multline}
Here we used the fact that $\E(X Y)=\E(X)\E(Y)$ if $X$ and $Y$ are independent variables and we note that $M_{q,\rho(q)}$ and $M^{*}_{\tau(q),\rho(q)}$ are independent of all other factors. We further use that the expectation of our i.i.d. complex Gaussian elements is zero, $\E(M_{ij})=0$ for all $ij$. If we again use the linearity of the expectation value and realize that $\prod^n_{i=1}\mathcal{S}_{i,\tau(i)}$ is completely independent of all elements in $M$, we find that 
\begin{multline}
    \E(Q_k)=\\
    \E\left( \sum_{j=k+1}^n\sum_{\tau \in \sigma_j} \left(\prod^n_{i=1}\mathcal{S}_{i,\tau(i)}\right) \text{Perm}(M \circ M^*_{\tau})\right)=\\ \sum_{j=k+1}^n\sum_{\tau \in \sigma_j} \left(\prod^n_{i=1}\mathcal{S}_{i,\tau(i)}\right)\E\left( \text{Perm}(M \circ M^*_{\tau})\right)=0,
\end{multline}

hence 
\begin{equation}
    \E(|Q_k|) \leq \sqrt{\E(Q_k^2)-\E(Q_k)^2} = \sqrt{\Var(Q_k)}.
\end{equation}
The square root of the variance thus provides an upper bound for $Q_k$. We derive an expression for $\Var(Q_k)$ in the remainder of this section. \\
Using Bienaymé's identity \cite{klenke_moments_2014} we find
\begin{multline}
    \Var(Q_k)\\
    =\Var\left(\sum_{j=k+1}^n\sum_{\tau \in \sigma_j}\left(\prod^n_{i=1}\mathcal{S}_{i,\tau(i)}\right) \Perm(M\circ M_{\tau}^*)\right) \\
    =\Var\left(\sum_{j=k+1}^n\sum_{\tau \in \sigma_j}\left(\prod^n_{i=1}\mathcal{S}_{i,\tau(i)}\right) \sum_{\rho\in\mathrm{S}_n}\prod_{r=1}^n(M\circ M_{\tau}^*)_{r,\rho(r)}\right)\\
    =\Var\left(\sum_{j=k+1}^n\sum_{\tau \in \sigma_j}\left(\prod^n_{i=1}\mathcal{S}_{i,\tau(i)}\right)\right.\times\\ 
    \left.\sum_{\rho\in\mathrm{S}_n}\prod_{r=1}^n\left(M_{r,\rho(r)} M_{\tau(r),\rho(r)}^*\right)\right)\\
    =\sum_{j=k+1}^n\sum_{j'=k+1}^n\sum_{\tau \in \sigma_j}\sum_{\tau' \in \sigma_{j'}} \sum_{\rho\in\mathrm{S}_n}\sum_{\rho'\in\mathrm{S}_n}\prod^n_{i=1}\mathcal{S}_{i,\tau(i)}\mathcal{S}^{*}_{i,\tau'(i)}\times\\
    \Cov\left(\prod_{r=1}^n\left(M_{r,\rho(r)} M_{\tau(r),\rho(r)}^*\right),\right.\cdots\\
    \left.\prod_{r=1}^n\left(M_{r,\rho'(r)} M_{\tau'(r),\rho'(r)}^*\right) \right).
    \label{eq: Var(Q) 6 sums}
\end{multline}
We use the definition of the covariance between two complex random variables to find
\begin{multline}
    \Cov\left(\prod_{r=1}^nM_{r,\rho(r)} M_{\tau(r),\rho(r)}^*,\prod_{r=1}^nM_{r,\rho'(r)} M_{\tau'(r),\rho'(r)}^*\right)\\
    =\E\left(\prod_{r=1}^n M_{r,\rho(r)} M_{\tau(r),\rho(r)}^* M^{*}_{r,\rho'(r)} M_{\tau'(r),\rho'(r)}\right)\cdots\\
    -\E\left(\prod_{r=1}^nM_{r,\rho(r)} M_{\tau(r),\rho(r)}^*\right)\times\\\E\left(\prod_{r=1}^nM^{*}_{r,\rho'(r)} M_{\tau'(r),\rho'(r)}\right)
    \label{eq: cov=}
\end{multline}
If we focus on the second term in Eq. (\ref{eq: cov=}), we realize that again, for all $\tau$ that we will consider, this term evaluates to zero for the same reasons as given in Eqs. (\ref{eq: at least 1 nonfixed point}) and (\ref{eq: exp Q=0 because at least 1 nonfixed point}). Then Eq. (\ref{eq: cov=}) reduces to 
\begin{multline}
    \Cov\left(\prod_{r=1}^nM_{r,\rho(r)} M_{\tau(r),\rho(r)}^*,\prod_{r=1}^nM_{r,\rho'(r)} M_{\tau'(r),\rho'(r)}^*\right)\\
    =\E\left(\prod_{r=1}^nM_{r,\rho(r)} M_{\tau(r),\rho(r)}^*M^{*}_{r,\rho'(r)} M_{\tau'(r),\rho'(r)}\right).
    \label{eq: cov= E[prod]  -0}
\end{multline}
We will continue to show that the expression in Eq. (\ref{eq: cov= E[prod]  -0}) is equal to zero for almost all of the combinations of $\tau$, $\rho$, $\tau'$ and $\rho'$. In this demonstration, it is crucial to assume that all elements of our submatrix $M$ are i.i.d. complex Gauassian, or:
\begin{equation}
    M_{ij}\sim \mathcal{CN}(0,\frac{1}{m})\:\forall\:i,j
    \label{eq: distributed}
\end{equation}
From Eq. (\ref{eq: distributed}) it follows that 
\begin{enumerate}
    \item $\E\left(M_{ij}\right)=0$
    \item $\E\left((M_{ij})^2\right)=\E\left((M^{*}_{ij})^2\right)=0$
    \item $\E\left(|M_{ij}|^2\right)=\frac{\E(\chi^2_2)}{2m}=\frac{1}{m}$
    \item $\E\left(|M_{ij}|^4\right)=\frac{2}{m^2}$
\end{enumerate}
If we inspect the equations listed above, we note that Eq. (\ref{eq: cov= E[prod]  -0}) will only evaluate to a nonzero amount when for all $r$ one of the following conditions is true
\begin{enumerate}
\item $r=\tau(r)$ and $r=\tau'(r)$
\item $\tau(r)=\tau'(r)$ and $\rho(r)=\rho'(r)$
\end{enumerate}
We note that the first condition is true for all $r\in\mathrm{fix}(\tau)$, if $\tau$ and $\tau'$ share the same fixed points. For all non-fixed points, condition 2 must thus be true and we conclude that only if $\tau=\tau'$ and $\rho(q)=\rho'(q)\: \forall\:q\in \mathrm{nonfix}(\tau)$, Eq. (\ref{eq: cov= E[prod]  -0}) may evaluate to a non-zero value. Here $\mathrm{fix}(\tau)$ and $\mathrm{nonfix}(\tau)$ denote the set of all fixed points of the permutation $\tau$ and its complementary set respectively.
Then, Eq.(\ref{eq: Var(Q) 6 sums}) reduces to:
\begin{multline}
    \Var(Q_k)=\\
    \sum_{j=k+1}^n\sum_{\tau \in \sigma_j} \sum_{\rho\in\mathrm{S}_n}\sum_{\substack{\rho'\in\mathrm{S}_n\\\rho(q)=\rho'(q)\: \\\forall\:q\in \mathrm{nonfix}(\tau)}}\prod^n_{i=1}|\mathcal{S}_{i,\tau(i)}|^2\times\\
    \E\left(\prod_{r=1}^n(M_{r,\rho(r)} M_{\tau(r),\rho(r)}^*)(M^{*}_{r,\rho'(r)} M_{\tau(r),\rho'(r)})\right).
\end{multline}

We now split up the product.
\begin{multline}
    \Var(Q_k)=\\
    \sum_{j=k+1}^n\sum_{\tau \in \sigma_j} \sum_{\rho\in\mathrm{S}_n}\sum_{\substack{\rho'\in\mathrm{S}_n\\\rho(q)=\rho'(q)\:\\ \forall\:q\in \mathrm{nonfix}(\tau)}}\prod^n_{i=1}|\mathcal{S}_{i,\tau(i)}|^2\times\\
    \E\left(\prod_{r\in\mathrm{fix}(\tau)}M_{r,\rho(r)} M_{\tau(r),\rho(r)}^*M^{*}_{r,\rho'(r)} M_{\tau(r),\rho'(r)}\right)\times\\\E\left(\prod_{q\in\mathrm{nonfix}(\tau)}M_{q,\rho(q)} M_{\tau(q),\rho(q)}^*M^{*}_{q,\rho'(r)} M_{\tau(q),\rho'(q)}\right)
\end{multline}
which, if we use that $\tau(r)=r\: \forall\:r\in \mathrm{fix}(\tau)$ and $\rho(q)=\rho'(q)\: \forall\:q\in \mathrm{nonfix}(\tau)$, reduces to
\begin{multline}
    \Var(Q_k)=\\
    \sum_{j=k+1}^n\sum_{\tau \in \sigma_j} \sum_{\rho\in\mathrm{S}_n}\sum_{\substack{\rho'\in\mathrm{S}_n\\\rho(q)=\rho'(q)\: \\\forall\:q\in \mathrm{nonfix}(\tau)}}\prod^n_{i=1}|\mathcal{S}_{i,\tau(i)}|^2\times\\
    \E\left(\prod_{r\in\mathrm{fix}(\tau)}|M_{r,\rho(r)}|^2 |M_{r,\rho'(r)}|^2 \right)\times\\\E\left(\prod_{q\in\mathrm{nonfix}(\tau)}|M_{q,\rho(q)}|^2 |M_{\tau(q),\rho(q)}|^2\right).
    \label{eq: var (Q) 4 sums}
\end{multline}
If we now realize that $M_{q,\rho(q)}$ is independent of $M_{\tau(q),\rho(q)}$ for all $q\in \mathrm{nonfix}(\tau)$, $M_{r,\rho(r)}$ is equal to $M_{r,\rho'(r)}$ if $\rho'(r)=\rho(r)$ and independent otherwise, $\E\left(|M_{ij}|^2\right)=\frac{1}{m}$ and $\E\left(|M_{ij}|^4\right)=\frac{2}{m^2}$, Eq. (\ref{eq: var (Q) 4 sums}) reduces to
\begin{multline}
    \Var(Q_k)=
    \sum_{j=k+1}^n \sum_{\tau\in\sigma_j}\prod^n_{i=1}|\mathcal{S}_{i,\tau(i)}|^2\sum_{\rho\in\mathrm{S}_n}\sum^{n-j}_{p=0}\cdots \\\mathrm{R}_{n-j,p}\left(\frac{1}{m^2}\right)^j\left(\frac{2}{m^2}\right)^p\left(\frac{1}{m^2}\right)^{n-j-p}\\
    =\sum_{j=k+1}^n \sum_{\tau\in\sigma_j}\prod^n_{i=1}|\mathcal{S}_{i,\tau(i)}|^2 n!\sum^{n-j}_{p=0}\mathrm{R}_{n-j,p}2^p\left(\frac{1}{m^2}\right)^{n}.
 %
    \label{eq: var (Q) intermediata}
\end{multline}
Here $R_{n,k}$ is rencontres number that counts the number of ways one can permute the set $\{1,\cdots, n\}$ with $k$ fixed points. Note that the number of permutation in the set $\sigma_{j}$ is given by $\mathrm{R}_{n,n-j}=\frac{n!}{j!}\sum_{q=0}^{n-j}\frac{(-1)^q}{q!}$. 

In general, we believe it is difficult to simplify Eq. (\ref{eq: var (Q) intermediata}) further. We therefore restrict ourselves to a generalized version of the orthogonal bad-bit model \cite{sparrow2018phd_thesis,marshall2022distillation}. Note that this model covers both the special cases where the overlap of all photons are identical, as well as the case where a subset of all photons is perfectly identical. In this model we consider the internal mode of the $i^{th}$ photon to be:
\begin{equation}
    |\psi\rangle = \sqrt{x_i}|\Psi_0\rangle + \sqrt{1-x_i}|\Psi_i\rangle,
    \label{eq: iid obb}
\end{equation}
where $\langle\Psi_i|\Psi_j\rangle=\delta_{ij}$. Every permutaion $\tau$ can uniquely be described by it's cycle notation. From the cycle notation, it becomes clear that for this model, $\sum_{\tau\in\sigma_j}\prod^n_{i=1}|\mathcal{S}_{i,\tau(i)}|^2=\mathrm{R}_{n,n-j} \mathrm{M}_j$. Where $\mathrm{M}_j$ is the elementary symmetric polynomial of order $j$ for the variables $|x_i|^2$, divided by $\begin{pmatrix}n\\j\end{pmatrix}$. We can use McLaurin's inequality to find an upper bound for the expression in Eq. (\ref{eq: var (Q) intermediata}).
\begin{equation}
\begin{split}
    \sum_{\tau\in\sigma_j}\prod^n_{i=1}|\mathcal{S}_{i,\tau(i)}|^2&=\mathrm{R}_{n,n-j} \mathrm{M}_j\\
    &\leq \mathrm{R}_{n,n-j} \left(\sqrt{\mathrm{M}_2}\right)^j
    \end{split}
    \label{eq: mclaurin's inequality}
\end{equation}
Where $\sqrt{\mathrm{M}_2}$ represents the square root of the second order elementary symmetric mean of $|x_i|^2$. Conveniently, this equals the quadratic mean of the pairwise HOM visibilities.

We can now simplify Eq. (\ref{eq: var (Q) intermediata}) as follows:
\begin{multline}
\Var(Q_k)\leq\\
    \\\sum_{j=k+1}^n \mathrm{R}_{n,n-j} \left(\sqrt{\mathrm{M}_2}\right)^{j} n!\sum^{n-j}_{p=0}\mathrm{R}_{n-j,p}2^p\left(\frac{1}{m^2}\right)^{n}=\\
    \sum_{j=k+1}^n \left(\sqrt{\mathrm{M}_2}\right)^{j} \frac{(n!)^2}{m^{2n}}\sum_{q=0}^{j}\frac{(-1)^q}{q!}\sum^{n-j}_{p=0}\frac{2^p}{p!}\sum_{r=0}^{n-j-p}\frac{(-1)^r}{r!}.
    \label{eq: var (Q) intermediatc}
\end{multline}
Now $\sum_{q=0}^{j}\frac{(-1)^q}{q!}\approx \frac{1}{e}$, 
$\sum^{n-j}_{p=0}\frac{2^p}{p!}\approx e^2$ and 
$\sum_{r=0}^{n-j-p}\frac{(-1)^r}{r!}\approx \frac{1}{e}$ when $j$, $n-j-p$ and $n-j$ are large respectively. And thus $\sum_{q=0}^{j}\frac{(-1)^q}{q!}\sum^{n-j}_{p=0}\frac{2^p}{p!}\sum_{r=0}^{n-j-p}\frac{(-1)^r}{r!}\approx 1$. 
\footnote{Note that for $j=4$, $n-j=5$ and $n-j-p=4$ these approximations already have errors below $2\%$.}
Hence,
\begin{equation}
    \Var(Q_k)\approx \sum_{j=k+1}^n \left(\sqrt{\mathrm{M}_2}\right)^{j} \frac{(n!)^2}{m^{2n}}.
    \label{eq: var (Q) final}
\end{equation}
We realize that Eq. (\ref{eq: var (Q) final}) describes a truncated geometric series.
\begin{equation}
\begin{split}
        \Var(Q_k)&\approx \frac{(n!)^2}{m^{2n}}\frac{(\sqrt{\mathrm{M}_2})^{k+1}-(\sqrt{\mathrm{M}_2})^{n+1}}{1-\mathrm{M}_1}\\
    &<\frac{(n!)^2}{m^{2n}}\frac{(\sqrt{\mathrm{M}_2})^{k+1}}{1-\sqrt{\mathrm{M}_2}}
\end{split}
\label{eq: final Var after geometric sum}
\end{equation}
Finally, we use Eq. (\ref{eq: final Var after geometric sum}) to find an upper bound for the expectation value of the $L_1$-distance between the approximate distribution and the real distribution over the Haar-unitaries. In the following expression the sum with index $s$, runs over all non-collisional outputs.
\begin{equation}
\begin{split}
    \E\left(\sum_s|P(s)-P_k(s)|\right)&=\E\left(\sum_s|Q_k(s)|\right)\\
    &=\sum_s\E\left(|Q_k(s)|\right)\\
    &<\begin{pmatrix}m\\n\end{pmatrix}\frac{n!}{m^n}\sqrt{\frac{(\sqrt{\mathrm{M}_2})^{k+1}}{1-\sqrt{\mathrm{M}_2}}}\\
    &<\sqrt{\frac{(\sqrt{\mathrm{M}_2})^{k+1}}{1-\sqrt{\mathrm{M}_2}}}
\end{split}
\label{eq: upper bound appendix}
\end{equation}


\bibliographystyle{apsrev}

\bibliography{refs.bib}

\end{document}